\documentclass[aps,prl,twocolumn,showpacs]{revtex4-1}
\usepackage[T1]{fontenc}
\usepackage{graphicx}
\usepackage{color}
\usepackage{amsmath}

\newlength{\halfcolumnwidth}
\begin{document}
\title{
Probing three-body correlations in a quantum gas  using the measurement of the third
moment of density fluctuations}
\date{\today}
\author{J.~Armijo$^{(1)}$, T.~Jacqmin$^{(1)}$, K.~V.~Kheruntsyan$^{(2)}$, and I.~Bouchoule$^{(1)}$}
\affiliation{$^{(1)}$Laboratoire Charles Fabry, UMR 8501 du CNRS,  Institut
d'Optique, 91 127 Palaiseau Cedex, France\\
$^{(2)}$ARC Centre of Excellence for Quantum-Atom Optics, School of Mathematics and Physics,
University of Queensland, Brisbane, Queensland 4072, Australia
}

\begin{abstract}
We perform measurements of the third moment of atom number fluctuations in small slices of
a very elongated weakly interacting
degenerate Bose gas. We find a positive skewness of the
atom number distribution in the ideal gas regime
and a reduced skewness compatible with zero in the quasi-condensate regime.
For our parameters, the third moment is a thermodynamic quantity whose
measurement constitutes a sensitive test of the equation of state
and our results are in agreement with a modified Yang-Yang
thermodynamic prediction. Moreover, we show that the measured skewness
reveals the presence of true three body correlations in the system.
\end{abstract}


\maketitle

Measurements of higher-order
correlations and the density fluctuations, in particular,
 are becoming an
increasingly important tool in the studies of ultracold quantum gases.
Such measurements are able to probe quantum many-body states of
 interacting
systems, often giving access to key quantities that
characterize the system.
This is particularly true for one-dimensional (1D) gases, where the effects of
 fluctuations
are enhanced compared to 3D systems and govern the rich underlying physics.
Zero-distance (or local) second- and third-order correlation functions
have been probed in several
ultracold gas experiments by measuring the
rates of two- and three-body inelastic
processes such as photoassociation and three-body
recombination\cite{PhysRevLett.79.337,PhysRevLett.92.190401,PhysRevLett.95.190406}.
Such measurements
enabled the study of the strongly correlated regime of `fermionization'
in a 1D Bose gas.

An alternative experimental technique is the \textit{in situ} measurement
of atom number fluctuations in a small detection volume, achievable
through the analysis of noise in absorption images.
The fluctuation variance (or second moment)
provides information about an integrated non-local
density-density correlation function.
In addition, under adequate experimental conditions,
the variance can render as a thermodynamic quantity
and therefore such measurements
constitute a
probe of the thermodynamic  
properties of quantum gases~\cite{esteve:130403,Hung2010,Armijo}, alternative 
to the analysis of density profiles~\cite{PhysRevLett.100.090402,Rath2010,Salomon2010} 
or momentum distributions in the context of Tan's relations \cite{JILA-thermodynamics}.
\textit{In situ} measurements of atom number fluctuations in weakly interacting
quasi-1D Bose gases were used to probe the crossover from the nearly ideal
gas regime, where bosonic bunching is present, to the quasi-condensate
regime, where
the density fluctuations are suppressed~\cite{esteve:130403,Armijo}.
In fermionic systems, sub shot-noise atom number fluctuations were observed in
a degenerate Fermi gas~\cite{Muller2010,Sanner2010}.
Combining this with the measurement
of compressibility of the gas deduced from the known density profile
and confining potential,
such measurements have been shown
to provide reliable thermometry~\cite{Muller2010}.

\begin{figure}[t]
\includegraphics[width=0.97\columnwidth]{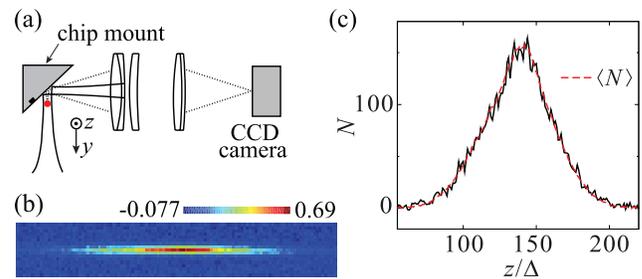}
\caption{(Color online) (a) Scheme of the imaging setup.
The probe laser 
crosses the atomic cloud 
(red dot) before its reflection
from the chip surface and detection on a CCD camera.
(b) Typical \textit{in-situ} absorption
image; the scale on the colorbar shows the optical density.
The pixel size in the object plane is $\Delta\!=\!4.5$~$\mu$m.
(c) Typical longitudinal density profile (solid curve), together with the
mean profile (dashed red curve).}
\label{fig.setup-and-images}
\end{figure}

In this paper we expand the arsenal of probes of higher-order
correlations in quantum gases by measuring the third moment
of atom number fluctuations. This is done using \textit{in situ}
absorption imaging of an ultracold gas on an atom-chip setup sketched in
Fig.~\ref{fig.setup-and-images}~(a).
We probe a weakly interacting quasi-1D Bose gas.
We have measured a positive
third moment of the atom number distribution in a degenerate gas
within the ideal gas regime and within the crossover towards a quasi-condensate.
In the quasi-condensate
regime the measured third moment is compatible with zero.
The third moment of the atom number distribution
is linked to the third-order correlation function and
our measurements demonstrate the presence of true three-body
correlations in the gas. Apart from this, we show that the measured third moment is
related to a thermodynamic relation that involves a second-order derivative, and therefore
the technique can be used as a sensitive probe of the thermodynamics of a quantum gas.

Our quasi-1D Bose gases are produced using $^{87}$Rb atoms
in the hyperfine state $|F\!=\!2,m\!=\!2\rangle$.
A very elongated Ioffe magnetic trap with a longitudinal
oscillation frequency ranging from $5.0$~Hz to $8$~Hz
and a transverse oscillation frequency
$\omega_\perp/2\pi$ ranging from $3$~kHz to $4$~kHz
is realized using on-chip micro-wires
and an external homogeneous magnetic field.
Using forced rf evaporation, we produce ultracold clouds at temperatures
from $T=20$~nK to $500$~nK.
The longitudinal rms size $L$ of the cloud ranges from $\sim\!\!50$~$\mu$m to
$\sim\!\!100$~$\mu$m.
As shown previously~\cite{esteve:130403}, under these conditions
such gases explore the crossover from the ideal gas
regime to the quasi-condensate regime, and the underlying physics
lies in the 1D regime or
in the crossover from 1D to 3D~\cite{Armijo}.

\textit{In situ} measurements of density fluctuations are performed using
absorption images such as the one shown
in Fig.~\ref{fig.setup-and-images}~(b). The details of our imaging  and
calibration techniques are described in EPAPS
Ref.~\cite{appendix}.
As the transverse size of the trapped cloud ($<500$~nm rms)
is much smaller than the pixel size (4.5~$\mu$m), the only information
in the transverse direction is the diffractional and motional blur on the image.
By summing the atom number over transverse pixels, we reduce the notion of a
pixel to a segment of length $\Delta$ and derive from each image the longitudinal density
profile [Fig.~\ref{fig.setup-and-images}~(c)].
We perform a statistical analysis of hundreds
of images taken under the same
experimental conditions~\cite{esteve:130403,Armijo}.
For each profile and pixel we extract $\delta N=N-\langle N\rangle$,
where $\langle N\rangle$
is given by the average density profile.
To remove the
effect of shot-to-shot
variations in the total
atom number $N_{\rm tot}$, the profiles are ordered
according to $N_{\rm tot}$ and
we use a running average of about 20 profiles.
As will be explained below, the longitudinal confining potential is
irrelevant and each $\delta N$ is binned according to
the corresponding mean atom number in the pixel $\langle N\rangle$.
For each bin,
we compute the second and third moment
of atom number fluctuations,
$\langle \delta N^{2} \rangle$ and $\langle \delta N^{3} \rangle$.
The contribution of the optical shot noise
to these quantities is subtracted, although it is negligible for $\langle \delta N^{3} \rangle$.

\begin{figure}[t]
\includegraphics[width=0.99\columnwidth]{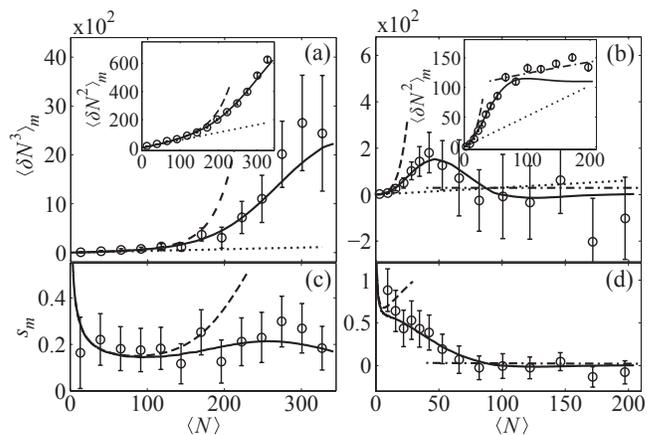}
\caption{Measured third moment (open circles) of the
atom number fluctuations versus the mean atom number per pixel, for temperatures of $376$~nK (a) and
$96$~nK (b). The insets show the corresponding atom number variances.
The errorbars are the statistical errors.
Graphs (c) and (d) show the skewness $s_{m}$
corresponding to (a) and (b), respectively.
The theoretical predictions, scaled by
$\kappa_2=0.55$ and $\kappa_3=0.34 $ for (a) and (c), and by
$\kappa_2=0.52$ and $\kappa_3=0.31 $ for (b) and (d), are shown for comparison: solid lines -- the modified Yang-Yang prediction; dashed lines -- the ideal Bose gas prediction; dash-dotted lines on (b) and (d) -- the quasi-condensate prediction; dotted lines -- the
shot noise limit $\langle N\rangle$.
}
\label{fig.deltaN3}
\end{figure}

The measured third moment of the atom number fluctuations,
$\langle \delta N^{3} \rangle _{m}$,
is plotted in
Fig.~\ref{fig.deltaN3} for two different temperatures.
For the higher temperature [Fig.~\ref{fig.deltaN3}~(a)],
we observe a positive value of
$\langle \delta N^3\rangle_m$ that increases with the average atom number $\langle N \rangle$.
 At a smaller temperature [Fig.~\ref{fig.deltaN3}~(b)],
$\langle \delta N^3\rangle_m$ initially grows with $\langle N \rangle$ and reaches a maximum,
before taking a value compatible with zero at large $\langle N \rangle$.
The corresponding second moments or variances $\langle \delta N^{2} \rangle_m$ are shown in the insets.
A finite third moment indicates an asymmetry of the atom number distribution,
which is usually quantified by the skewness of the distribution,
$s_m=\langle \delta N^3\rangle_m/\langle \delta N^2\rangle_m^{3/2}$, shown
in Figs.~\ref{fig.deltaN3} (c) and (d). Before discussing the physics and theoretical
understanding of these results, we first describe how the \textit{measured}
moments $\langle \delta N^{3} \rangle_m$ and $\langle \delta N^{2} \rangle_m$ are related
to the \textit{true}
moments $\langle \delta N^{3} \rangle$ and $\langle \delta N^{2} \rangle$.

The measurements of atom number fluctuations are affected by
the finite spatial resolution due to
both the optical resolution and the
diffusion of atoms during the optical pulse,
which cause the absorption signal from each atom to spread over
several pixels and blur the image.
Denoting by ${\cal A}$ the
impulse response function of the imaging system, the
impulse response for the pixel $[0,\Delta]$
is ${\cal F}(z_0)=\int_0^\Delta dz {\cal A}(z-z_0)$, and  the
measured atom number fluctuation in the pixel is given by
$\delta N_m=\int_{-\infty}^{+\infty} dz_0
{\cal F}(z_0) \delta n(z_0)$,
where $\delta n(z_0)$ is the local density fluctuation.
For the parameters explored in this paper,
the expected correlation length $l_c$ of density
fluctuations~\cite{PhysRevA.79.043619} is smaller than 0.5~$\mu$m. This is
sufficiently
smaller than the width of ${\cal A}$ so that
we can assume that the density fluctuations have
zero range. Moreover,
since the resolution and the pixel size are much smaller
than the longitudinal size of the atomic cloud, we can assume
that the gas is locally homogeneous with respect to $z$.
Then, the \textit{measured} second and third moments of
density fluctuations can be obtained as
\begin{align}
\langle\delta N^{2}\rangle_{m}  & =\langle\delta N^{2}\rangle\textstyle\int
_{-\infty}^{+\infty} \!{dz_{0}\;\mathcal{F}(z_{0})^{2}}/{\Delta}=\kappa_{2}\langle\delta
N^{2}\rangle,\label{eq.kappa2}\\
\langle\delta N^{3}\rangle_{m}  & =\langle\delta N^{3}\rangle\textstyle\int
_{-\infty}^{+\infty} \!{dz_{0}\;\mathcal{F}(z_{0})^{3}}/{\Delta}=\kappa_{3}\langle\delta
N^{3}\rangle,\label{eq.kappa3}%
\end{align}
where $\langle \delta N^2\rangle$ and $\langle\delta N^3\rangle$ are the
respective \textit{true} moments, whereas $\kappa_{2}$ and $\kappa_{3}$ are
the reduction factors.
For low enough linear densities, the gas lies in the
nondegenerate ideal gas regime. Then the fluctuations are
almost that of a Poissonian
distribution, so that  $\langle \delta N^2\rangle\simeq
\langle \delta N^3\rangle\simeq\langle N\rangle$, and
 the reduction factors
may be deduced from a linear fit of the measured fluctuations versus
$\langle N\rangle$, where $\langle N\rangle$
is experimentally determined absolutely.
However, such a deduction is difficult
in very cold clouds where only few pixels lie in the
nondegenerate ideal gas regime.

We thus develop an alternative method that
uses the measurement of the atom number
correlation 
\begin{equation}
C_{i,i+j}\!=\!\frac{\langle \delta N_{i}\delta N_{i+j}\rangle _{m}}{\langle \delta
N_{i}^{2}\rangle _{m}}\!=\!\frac{\int_{-\infty}^{+\infty} \!dz_{0}\mathcal{F}(z_{0})\mathcal{F}
(z_{0}\!-\!j\Delta )}{\int_{-\infty}^{+\infty} \!dz_{0}\;\mathcal{F}(z_{0})^{2}}
\label{eq.cor}
\end{equation}%
between the pixel $i$ and the adjacent ($j\!=\!1$) or
the next-neighbor ($j\!=\!2$) pixels.
Such correlation arises due to the contribution of an atom to the absorption
in both pixels.
Making a Gaussian ansatz for the impulse
response function ${\cal A}$,
the rms width of ${\cal A}$ can obtained by fitting Eq.~(\ref{eq.cor}) to
the measured correlations $C_{i,i+1}$ \cite{appendix}.
The reduction factors $\kappa_{2,3}$ 
can then be deduced from Eqs.~(\ref{eq.kappa2}) and (\ref{eq.kappa3}),
resulting typically in 
$\kappa_3 \simeq 0.3$ and $\kappa_2 \simeq 0.5$ for our data.
The result 
is in good
agreement with the
slope of $\langle \delta N^2\rangle_m$ at small $\langle N\rangle$
[see the inset of Fig.~\ref{fig.deltaN3}~(a)].

Turning to the discussion of the physics behind our experimental results, we
first point out that the third moment of atom number fluctuations is actually
a thermodynamic quantity when, as in our experiment,
the pixel size is both much larger than the characteristic correlation length of density fluctuations
$l_c$ and much smaller than the
cloud length $L$, $l_c\!\ll\! \Delta \!\ll \!L$.
Then a local density approximation is valid and
the gas contained in a
pixel can be well described by a grand canonical ensemble,
in which the rest of the cloud is acting as a reservoir that fixes
the chemical potential
$\mu$ and the temperature  $T$.
Denoting by $\mathcal{Z}$ the grand-canonical partition
function, we have $\langle N\rangle\! =\!(k_{B}T/\mathcal{Z})\partial \mathcal{Z}
/\partial \mu $, $\langle N^{2}\rangle\! =\!(k_{B}^{2}T^{2}/\mathcal{Z})\partial ^{2}
\mathcal{Z}/\partial \mu ^{2}$ and $\langle N^{3}\rangle \!=\!(k_{B}^3T^3/\mathcal{Z})
\partial ^{3}\mathcal{Z}/\partial \mu ^{3}$. From the first two equations,
we obtain the well known thermodynamics relation $\langle \delta N^{2}\rangle \!=\!k_{B}T\partial
\langle N\rangle /\partial \mu $, whereas the three equations give the following relation
\begin{equation}
\langle \delta N^3\rangle =\left (k_BT\right )^2\partial^2\langle N\rangle/\partial \mu^2,
\label{eq.deltaN3}
\end{equation}
where $\langle N\rangle=n\Delta$, and $n$ is the linear
density of a gas homogeneous along $z$.
Thus, the knowledge of the equation of state 
(EoS) $n=n(\mu,T)$ is sufficient to
predict the third moment of the atom number distribution.
Note that a more traditional 
form of the EoS for pressure $P$ can be readily 
deduced from $n(\mu,T)$ using Gibbs-Duhem relation, $n\!=\!(\partial P / \partial \mu)_{T}$, 
leading to $P\!=\!\int _{-\infty} ^{\mu} n(\mu^{\prime},T) d\mu^{\prime}$.

We now compare our measurements with the predictions from
different models for the EoS $n(\mu,T)$.
The temperature of the cloud for the case of Fig.~\ref{fig.deltaN3}~(a)
is deduced from an ideal Bose gas fit to the wings of the density profile~\cite{Armijo}.
For the data of Fig.~\ref{fig.deltaN3}~(b), corresponding to the quasi-condensate regime, such wings
are vanishingly small and hard to detect.
In this case we deduce the temperature~\cite{Armijo}
from the measurement of density fluctuations in the cloud centre
using the thermodynamic relation $\langle \delta N^{2}\rangle \!=\!k_{B}^2T^2\partial
 \langle N\rangle /\partial \mu $ and the EoS of a
quasi-condensate (see below).

The predictions from the equation
of state for an ideal Bose gas are shown by the dashed
lines in Fig.~\ref{fig.deltaN3}. For a  highly
 nondegenerate (or classical) gas,
corresponding to small $\langle  N\rangle$,
this model predicts
$\langle \delta N^3\rangle\simeq
\langle \delta N^2\rangle \simeq\langle  N\rangle$ as
expected for a gas of uncorrelated particles.
When the gas becomes degenerate with the increase of $\langle  N\rangle$,
the contribution of the quantum-statistical exchange interaction term
to $\langle \delta N^3\rangle$ is no longer negligible, and
$\langle \delta N^3\rangle$ becomes
larger than the shot-noise term $\langle  N\rangle$.
Such an increase is observed in the experimental data
in Fig.~\ref{fig.deltaN3}~(a). However,
the ideal Bose gas model strongly overestimates the third moment
with further increase of $\langle  N\rangle$ and we eventually observe large discrepancy
between the predictions of this model and the experimental data.
The discrepancy is due to the repulsive interactions between the atoms, which
reduce the energetically costly density fluctuations.

Describing the effects of atomic interactions beyond the perturbative regime
is a challenging theoretical problem.
However, a purely 1D Bose gas with contact repulsive interactions is
a particular case of a many-body problem for which an exact solution is available through the
Yang-Yang thermodynamic formalism~\cite{Yang} in the entire parameter space.
For
the temperatures corresponding to Fig.~\ref{fig.deltaN3}~(a)-(b), the ratios of $k_{B}T/\hbar\omega_{\perp}$ are $2.6$
and $0.50$, respectively, implying that the population of the
transverse excited states is not negligible.
Accordingly, we use a \textit{modified} Yang-Yang model~\cite{PhysRevLett.100.090402},
in which the transverse ground state is described within the exact Yang-Yang
theory \cite{Isabelle-Karen-Gora},
 whereas the transverse excited states are treated as ideal 1D Bose gases.
This model has been shown to be valid for our parameters until the quasi-condensate
regime is reached~\cite{Armijo}. The corresponding predictions are plotted in Fig.~\ref{fig.deltaN3} and 
show that the model accounts
for the measured $\langle \delta N^3\rangle$ very well.

In the quasi-condensate regime [corresponding to
$\langle N\rangle\!\gtrsim\!70$
in Fig~\ref{fig.deltaN3}~(b)], where the density fluctuations are
suppressed~\cite{Quasibec_Castin,Kheru2003},
the EoS can be obtained numerically
from the 3D Gross-Pitaevskii equation
and is well described by the heuristic function
$\mu=\hbar\omega_\perp(\sqrt{1+4na}-1)$~\cite{PhysRevA.68.043610}.
In contrast to the modified Yang-Yang model,
this EoS
accounts for the transverse swelling of the cloud due to interatomic interactions
and better describes the measured
variance [see the inset of Fig.~\ref{fig.deltaN3}~(b)].
The measured third moment is compatible with this EoS.

\begin{figure}[t]
\includegraphics[width=0.99\columnwidth]{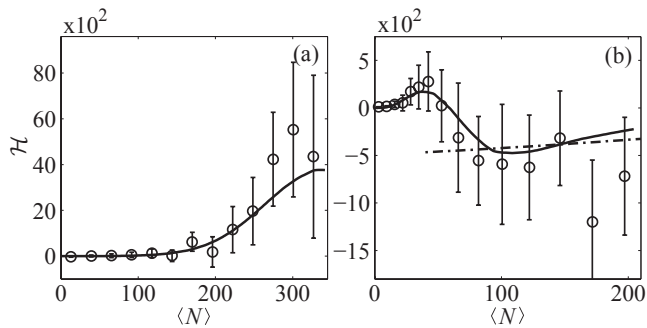}
\caption{The measured three-body integral ${\cal H}_{m}$ versus
$\langle N \rangle$, corresponding
to the data of Figs.~\ref{fig.deltaN3} (a) and (b), respectively.
The solid (dash-dotted) curves are the thermodynamic predictions
from the modified Yang-Yang (quasi-condensate) model.}
\label{fig.h}
\end{figure}

To unveil the role of many-body correlations, which underly the measured density
fluctuations while remaining hidden in the thermodynamic analysis, we consider the 1D two-
and three-body ($k\!=\!2,3$) correlation functions,
\begin{equation}
\tilde{g}^{(k)}(z_{1},\dots ,z_{k})\!=\!{\langle \tilde{\psi}^{\dagger }}
(z_{1})\dots \tilde{\psi}^{\dagger }(z_{k})\tilde{\psi}(z_{k})\dots
\tilde{\psi}(z_{1})\rangle /n^{k},
\end{equation}
where $\tilde{\psi}(z)=\!\int \!dxdy\;\psi (x,y,z)$ and $\psi $ ($\psi^{\dagger}$) is the bosonic
field annihilation (creation) operator.
 Using standard commutation relations
and the
 expression $\langle N^{2}\rangle =\langle N\rangle ^{2}+\langle N\rangle
 +n^{2}\iint\! dz_{1}dz_{2}[\tilde{g}^{(2)}({z_{1}},{z_{2}})-1]$, we find
\begin{eqnarray}
\langle \delta N^{3}\rangle  &=&\langle N\rangle +n^{3}\!\textstyle\iiint\!
dz_{1}dz_{2}dz_{3}\left[ \tilde{g}^{(3)}(z_{1},z_{2},z_{3})-1\right]
\nonumber \\
&&-3\langle N\rangle n^{2}\!\textstyle\iint\! dz_{1}dz_{2}[\tilde{g}%
^{(2)}(z_{1},z_{2})-1]  \nonumber \\
&&+3n^{2}\!\textstyle\iint\! dz_{1}dz_{2}[\tilde{g}^{(2)}(z_{1},z_{2})-1],
\label{eq.deltaN3g3}
\end{eqnarray}%
where the integrals are in the interval $[0,\Delta ]$.
As we see the third moment of atom number distribution depends
on both the $\tilde{g}^{(3)}$ and $\tilde{g}^{(2)}$ functions.
Moreover,
$\tilde{g}^{(3)}$ contains a contribution from $\tilde{g}^{(2)}$ since,
when one of the three particles is far from the other two,
$\tilde{g}^{(3)}$ reduces to $\tilde{g}^{(2)}$.
To remove such contributions,
we introduce the $h$-function
\begin{eqnarray}
h(z_{1},z_{2},z_{3}) \!\!&=&\!\!2+\tilde{g}^{(3)}(z_{1},z_{2},z_{3})  \label{eq.h} \\
&&-[\tilde{g}^{(2)}(z_{1},z_{2})\!+\!\tilde{g}^{(2)}(z_{2},z_{3})\!+\!\tilde{g}%
^{(2)}(z_{1},z_{3})], \notag
\end{eqnarray}
which is nonzero only for $z_1$, $z_2$ and $z_3$ being all in the
vicinity of each other.
Such a decomposition has been previously used
in the description of weakly correlated
plasmas~\cite{BBGKYplasma}, with the approximation $h\simeq 0$
being used to truncate
the
BBGKY hierarchy.

Using the $h$-function, Eq.~(\ref{eq.deltaN3g3}) can be rewritten as
\begin{eqnarray}
\langle \delta N^{3}\rangle  &=&\langle N\rangle +3n^{2}\!\textstyle\iint
\!dz_{1}dz_{2}[\tilde{g}^{(2)}(z_{1},z_{2})-1]  \notag \\
&& + \;n^{3}\!\textstyle\iiint \!dz_{1}dz_{2}dz_{3}h(z_{1},z_{2},z_{3}).
\end{eqnarray}%
Here, the first two terms represent one- and two-body effects,
with the second term being equal to $3\langle \delta N^2\rangle-3\langle N\rangle$.
The contribution of \textit{true} three-body correlations to
$\langle \delta N^3\rangle$ comes from the three-body integral
\begin{equation}
\mathcal{H}
\!=\!\!\langle \delta N^{3}\rangle +2\langle N\rangle -3\langle \delta N^{2}\rangle \!
\!=\!n^{3}\!\!\textstyle\iiint
\!dz_{1}dz_{2}dz_{3}h(z_{1},z_{2},z_{3}).  \label{eq.Hint}
\end{equation}%

In Fig.~\ref{fig.h}, we plot the
measured value of ${\cal H}$. More precisely, taking into account the
reduction factors $\kappa_2$ and $\kappa_3$, we plot
$\mathcal{H}=\langle \delta N^3\rangle_m/\kappa_3+2\langle N\rangle-
3\langle \delta N^2\rangle_m/\kappa_2 $.
We observe nonzero values of ${\cal H}$, which is a signature of
the presence of true three-body correlations in the gas:
${\cal H}$ is positive within the ideal gas regime
and in the crossover region towards the quasi-condensate [see Fig.~\ref{fig.h} (a)],
whereas it is 
negative in the quasi-condensate regime [Fig.~\ref{fig.h} (b)].
The results are in agreement with the thermodynamic predictions
of the modified Yang-Yang and the quasi-condensate models.

In summary, we have measured the 
third moment of density fluctuations 
in an ultracold quantum gas. 
This quantity
reveals the presence of true three-body correlations in the system.
Moreover, for sufficiently large pixels, 
such measurements constitute a very sensitive 
probe of the thermodynamics of the gas. 
As the third moment is related directly to the second-order 
derivative of the equation of state $n(\mu,T)$,
{the method 
lends itself as a high-precision tool for 
discriminating between alternative theoretical 
models, and can be applied to a broad class of ultracold atom systems.
For example, intriguing opportunities are in the understanding of the 
role of higher-order correlations 
in thermalisation 
of isolated quantum systems \cite{Thermalization}
and in the study of thermodynamics of more exotic 
many-body systems where three-body effects, such as Efimov 
resonances \cite{Efimov}, may lead to different signatures
in the second- and third-order correlations.

\begin{acknowledgments}
The Atom Optics group of the Laboratoire Charles-Fabry is a
member of the IFRAF Institute. This work was supported by
the ANR grant ANR-08-BLAN-0165-03 and by the Australian Research Council.
\end{acknowledgments}


\appendix
\section{Precise measurement of the atomic density profile and 
determination of imaging resolution}

A probe laser beam, locked onto the $D_2$ transition at
the wavelength $\lambda=780$~nm, is reflected from the chip
surface (covered by a gold mirror) after passing through the atomic cloud.
The shadow image of the atomic cloud is then recorded on a CCD camera,
with quantum efficiency larger than
90\%.
The diffraction-limited optical resolution has an rms width of  $1.0$~$\mu$m.
 Great care has to be taken in absorption imaging to allow for a reliable measurement of the atom number in each pixel.


We have chosen a configuration that maximizes the absorption efficiency. To achieve this, we focus
the probe beam onto the chip surface using a cylindrical lens as depicted in Fig.~1~(a) of the main text.
The $1/e^2$ size of the beam in the focused direction $x$ is $50~\mu$m,
which is smaller than the distance
of the atomic cloud from the chip surface so that the beam
crosses the atomic cloud only on its way to the chip and not
after its reflection.
Images as in Fig.~\ref{fig:images}~(a)-(c) are taken after
switching off the wire currents so that
only the external homogeneous magnetic field, whose  orientation
is close to the $y$ direction [see Fig.~1 (a) of the main text for the
axis definition], remains switched on.
With this geometry, using a $\sigma_+$-polarized probe beam we address only
the closed transition $|F=2,m=2\rangle \rightarrow |F'=3,m'=3\rangle$, and the
absorption cross-section at low intensity
takes its maximum value
$\sigma_0=3\lambda^2/2\pi$.

\begin{figure}[tb]
\includegraphics[width=0.95\columnwidth]{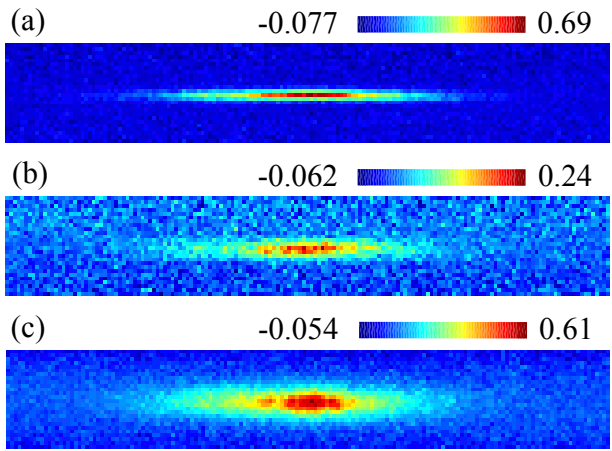}
\caption{(Color online) Typical absorption images used for calibration and the analysis of density fluctuations: (a) -- taken with a nearly
resonant probe, after a small time of flight of $0.5$~ms [same as Fig.~1 (b) of the main text, except with a larger field of view]; (b) -- taken with a probe detuned by
$5$~MHz and a time of flight of $1.5$~ms; (d) -- taken with a resonant probe and
a time of flight of $2.2$~ms to measure the total atom number.
The scales on the colorbars correspond to optical densities.}
\label{fig:images}
\end{figure}

The  absorption is measured by taking two pictures, the first one with the
atomic cloud present and the second one without the atoms.
The atom number in a given pixel $N_{p}$ is estimated from
the Beer-Lambert law $N_{p}=\ln(N_2/N_1)\Delta^2/\sigma $,
 where
$N_1$ and $N_2$ are the photon numbers in the pixel on the first
and second image, respectively, and $\sigma$ is the atomic
absorption cross-section.
With the transverse size of the cloud being smaller than
the pixel size, no information is available in the transverse direction,
and the number of atoms $N_{BL}$
in an effective pixel of size $\Delta$ can be
obtained by summing $N_{p}$ over
of the transverse pixels.
However, as the use of a resonant probe at high atomic densities produces
high optical densities (up to $1.5$) this naive procedure fails to correctly
estimate the true atom number $N$ in the pixel.
Firstly, when the transverse extension of the cloud is smaller than both
the pixel size and the optical resolution, the Beer-Lambert law underestimates
the true atom number due to the concavity of the logarithm function,
as already pointed out in \cite{esteve:130403epaps}.
Moreover, the validity of the Beer-Lambert law is questionable for
high atomic densities due to nontrivial reabsorption effects that may arise~\cite{Rath2010}.
 In order to reduce these effects, without decreasing too much the absorption,
we use a near resonant probe and enable the cloud to spread transversally during a
small time of flight of about $0.5$~ms -- sufficient to
reduce the effects of high atomic densities, but small enough
so that the atom number fluctuations in a pixel are barely affected.
The measured atom number $N_{BL}$, however,
still deviates from the true atom number $N$ and
we introduce a function $f$ defined as $N=f(N_{BL})$ to describe the
deviation from linearity at high optical densities.

\begin{figure}[tb]
\includegraphics[height=3.5cm]{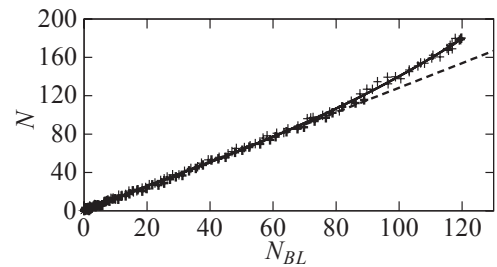}
\caption{Experimental determination of the correction to the Beer-Lambert law.
The solid curve is a fit to the experimental data with a
third-order polynomial,
the straight dashed line being the linear contribution.
$N_{BL}$ is obtained by using the
Beer-Lambert law and summing over transverse pixels in images
taken with a nearly resonant probe and a small time of flight of $0.5$~ms [see
Fig.~\ref{fig:images}~(a)].
$N$ is obtained from images taken with a probe detuned by
$5$~MHz and a time of flight of $1.5$~ms, for which the Beer-Lambert law is
adequate [see Fig.~\ref{fig:images}~(b)].}
\label{fig.functionf}
\end{figure}

The function $f$ is deduced from the comparison, in each effective pixel,
of the measured $N_{BL}$ with the
correct atom number $N$. The latter itself is measured as follows.
First, the correct profile is obtained using images [such as that
shown in Fig.~\ref{fig:images}~(b)]
taken with a
detuning of about $5$~MHz that reduces the absorption cross-section and a time of
flight of $\sim \!1.5$~ms that permits a transverse expansion of the cloud.
We checked experimentally that these parameters ensure the
validity of the Beer-Lambert law, while the
expansion is small enough as to retain the longitudinal profile
essentially unaffected.
Second, the absolute normalization
of the atomic density profile (or, equivalently, a measure of the
absorption cross-section $\sigma$),
is deduced form the knowledge of the total atom number.
 The latter is measured using a resonant probe with a
time of flight of $\sim \!2.5$~ms, as in Fig.~\ref{fig:images}~(c).
With such time of flight, the
cloud transverse expansion is sufficiently large and the
atomic density is small enough as to render the Beer-Lambert law applicable.
Atomic saturation is taken into account
via the formula $\sigma=\sigma_0(1+I/I_{\rm{sat}})$,
where $I$ is the intensity of the probe beam and $I_{sat}$ is the
saturation intensity.
A fit of
the measured absorption versus $I$ gives
$I_{\rm{sat}}=1.4(1)~$mW/cm$^{2}$, which is close to
the reported value of 1.62~mW/cm$^{2}$~\cite{SteckRbline}.
The remaining discrepancy could be because of possible underestimation
of the intensity of the probe beam seen by the atoms due to the losses during
the reflection of the beam
from the gold mirror and during the transmission through the optical lenses.
Finally, the function $f$ is  estimated by
fitting  the experimental points $N$ versus $N_{BL}$
with a third-order polynomial, as shown in Fig.~\ref{fig.functionf}.

All these calibrations are performed using values averaged over tens of
experimental realizations.
The images used for the analysis of fluctuations [as in
Fig.~\ref{fig:images}~(a)] and those used for calibration [as in
Figs.~\ref{fig:images}~(b) and~(c)] are taken in an alternated way -- typically
one picture of type (b) and then (c) after every 15 images of type (a) -- to
eliminate the dependence on noise arising from long-time magnetic field and thermal drifts.

The normalization procedure described above,
while compensating
for the effect of the small transverse size of the atomic cloud,
does not compensate for a possible error induced by
short-scale longitudinal density fluctuations.
However, in our experimental situation,
those fluctuations are smeared out by the atomic diffusion
during the probe pulse and are small.

\begin{figure}[tbh]
\includegraphics[height=3.5cm]{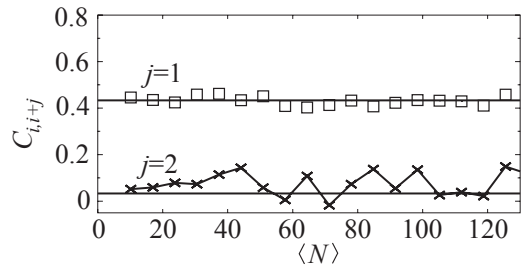}
\caption{Correlation of atom number fluctuations between adjacent
(open squares) and next-neighbor (crosses) pixels.
The solid lines are the predictions for an impulse response function
with an rms width of $2.0~\mu$m, obtained by fitting $C_{i,i+1}$.}
\label{fig.corr}
\end{figure}

The precise calibration of the atom number measurement described above is 
not sufficient for the analysis of atom number fluctuations. 
Indeed, the fluctuations are affected by a finite spatial resolution due
to
both the optical resolution and the
diffusion of atoms during the optical pulse,
which cause the absorption signal from each atom to spread over
several pixels and blur the image (see main text).
The finite  spatial resolution is also responsible for a nonvanishing 
correlation between the atom numbers measured in nearby pixels. 
In fact, we make use of this correlation for experimental determination of the spatial resolution of our imaging system.
 In Fig. \ref{fig.corr} we show the experimental data for 
the correlation coefficients $C_{i,i+j}$ [see Eq. (3) of the main text], 
corresponding to the atom number fluctuations in the adjacent ($j=1$) and 
next-neighboring ($j=2$) pixels, for the experimental data of Fig.~2~(c) of the 
main text. Fitting these correlation coefficients, we extract 
the rms width $\delta$ of the impulse response function ${\cal A}$ and we find that  
$\delta=2.0~\mu$m for this data set.

\end{document}